\documentclass[lettersize,journal]{IEEEtran}
\usepackage{amsmath,amsfonts}
\usepackage{algorithm}
\usepackage{array}
\usepackage{textcomp}
\usepackage{stfloats}
\usepackage{url}
\usepackage{verbatim}
\usepackage{booktabs}
\usepackage{xcolor}
\usepackage{makecell}
\usepackage{algorithm} 
\usepackage{algpseudocode}
\usepackage{amssymb}
\usepackage{hyperref}
\usepackage{multirow}
\usepackage{booktabs, tabularx}
\usepackage{pifont}
\usepackage{graphicx}
\usepackage[caption=false,font=footnotesize]{subfig}

\begin{document}

\title{Federated Learning and Unlearning for Recommendation with Personalized Data Sharing}

\author{
    Liang Qu$^{1}$~\IEEEmembership{Member,~IEEE},
    Jianxin Li$^{1\ast}$~\IEEEmembership{Senior Member,~IEEE},
    Wei Yuan$^{2}$,
    Shangfei Zheng$^{3}$~\IEEEmembership{Member,~IEEE},\\
    Lu Chen$^{4}$,
    Chengfei Liu$^{4}$~\IEEEmembership{Member,~IEEE}, and
    Hongzhi Yin$^{2}$~\IEEEmembership{Senior Member,~IEEE}%
    \thanks{$^{1}$School of Business and Law, Edith Cowan University, Perth, Australia.}%
    \thanks{$^{2}$School of Electrical Engineering and Computer Science, The University of Queensland, Brisbane, Australia.}%
    \thanks{$^{3}$School of Computer Science and Technology, Zhejiang Sci-Tech University, Hangzhou, China.}%
    \thanks{$^{4}$Department of Computing Technologies, Swinburne University of Technology, Hawthorn, VIC 3122, Australia.}%
    \thanks{$^{\ast}$Corresponding authors: Jianxin Li (jianxin.li@ecu.edu.au).}%
}

\markboth{Journal of \LaTeX\ Class Files,~Vol.~14, No.~8, August~2021}%
{Shell \MakeLowercase{\textit{et al.}}: A Sample Article Using IEEEtran.cls for IEEE Journals}


\maketitle

\begin{abstract}
Federated recommender systems have emerged as a paradigm for protecting user privacy by keeping interaction data on local devices while coordinating model training through a central server. However, most existing federated recommender systems adopt a one-size-fits-all assumption on user privacy, where all users are required to keep their data strictly local. This setting overlooks users who are willing to share their data with the server in exchange for better recommendation performance. Although several recent studies have explored personalized user data sharing in federated recommender systems, they assume static user privacy preferences and cannot handle user requests to remove previously shared data and its corresponding influence on the trained model. To address this limitation, we propose FedShare, a federated learn–unlearn framework for recommender systems with personalized user data sharing. FedShare not only allows users to control how much interaction data is shared with the server, but also supports data unsharing requests by efficiently removing the influence of the unshared data from the trained model. Specifically, in the learning phase, FedShare leverages shared data to construct a server-side high-order user–item graph and uses contrastive learning to jointly align local and global representations, enabling the model to capture both low-order and high-order collaborative signals. In the unlearning phase, we design a contrastive unlearning mechanism that selectively removes representations induced by the unshared data using a small number of historical embedding snapshots, avoiding the need to store large amounts of historical gradient information as required by existing federated recommendation unlearning methods. Extensive experiments on three public datasets demonstrate that FedShare achieves strong recommendation performance in both the learning and unlearning phases, while significantly reducing storage overhead in the unlearning phase compared with state-of-the-art baselines.
\end{abstract}


\section{Introduction}
Recommender systems \cite{qu2026sparse,long2025harnessing,qu2025efficient,zhang2025m2rec} have been widely used in many real-world applications to help users find what they may like from massive information. For example, video platforms, such as YouTube \cite{covington2016deep}, recommend videos that users may be interested in based on their watching history. Similarly, e-commerce platforms, such as Amazon \cite{smith2017two}, suggest products that users may want to buy according to their purchase history. Generally, most existing recommender systems rely on collecting user–item interaction data (e.g., watching history and purchase history) on a central server, and these collected data are then used to train recommendation models to learn users’ preferences for providing personalized recommendation services.
However, such centralized data collection may raise users’ concerns about data privacy, and may violate data protection regulations, such as the General Data Protection Regulation (GDPR\footnote{https://gdpr-info.eu/}). For example, Amazon was fined 746 million euros for GDPR violation\footnote{https://luxtoday.lu/en/business-economic-en/court-in-luxembourg-confirms-amazons-record-fine-746-million-euros-for-gdpr-violation}. 
Therefore, there exists a dilemma in which recommender systems inevitably need to collect users’ private interaction data for training models and providing personalized recommendation services.

\begin{figure*}[t]
  \centering
  \subfloat[Architecture of federated recommender system.\label{fig:intro_a}]{
    \includegraphics[width=0.48\textwidth]{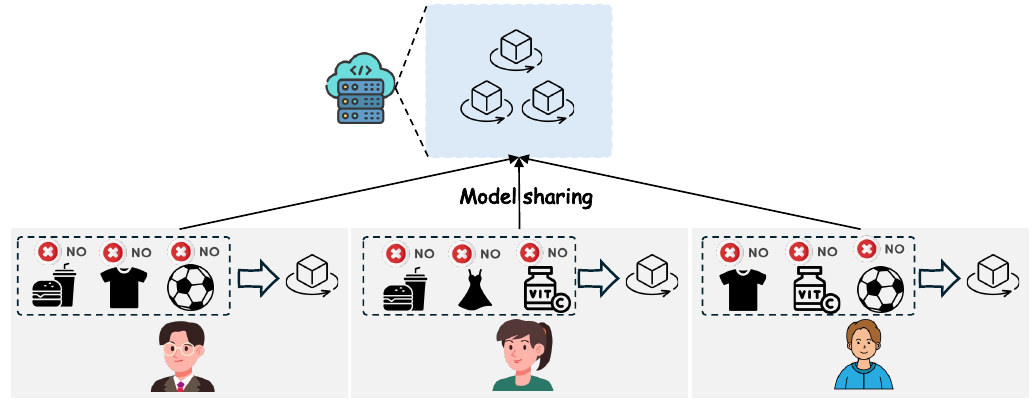}
  }\hfill%
  \subfloat[Federated recommender systems with personalized user data sharing.\label{fig:intro_b}]{
    \includegraphics[width=0.48\textwidth]{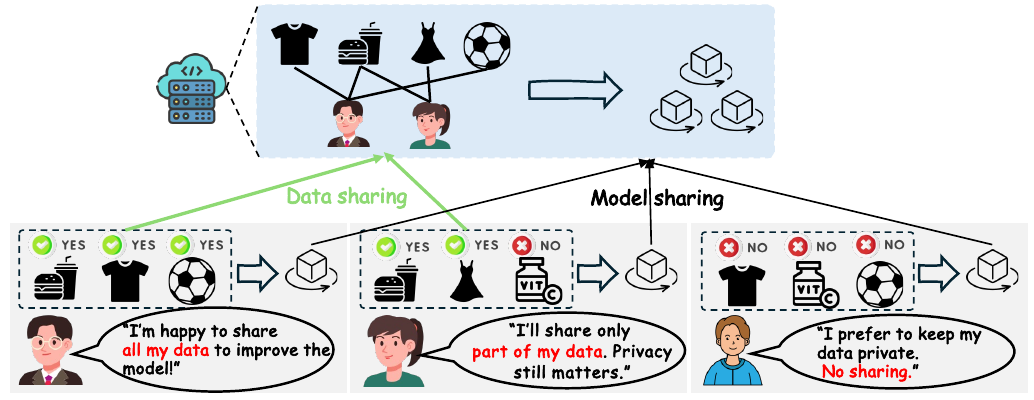}
  }

  \vspace{0.6em}

  \subfloat[Federated recommender systems with personalized user data sharing and unlearning support.\label{fig:intro_c}]{
    \includegraphics[width=0.9\textwidth]{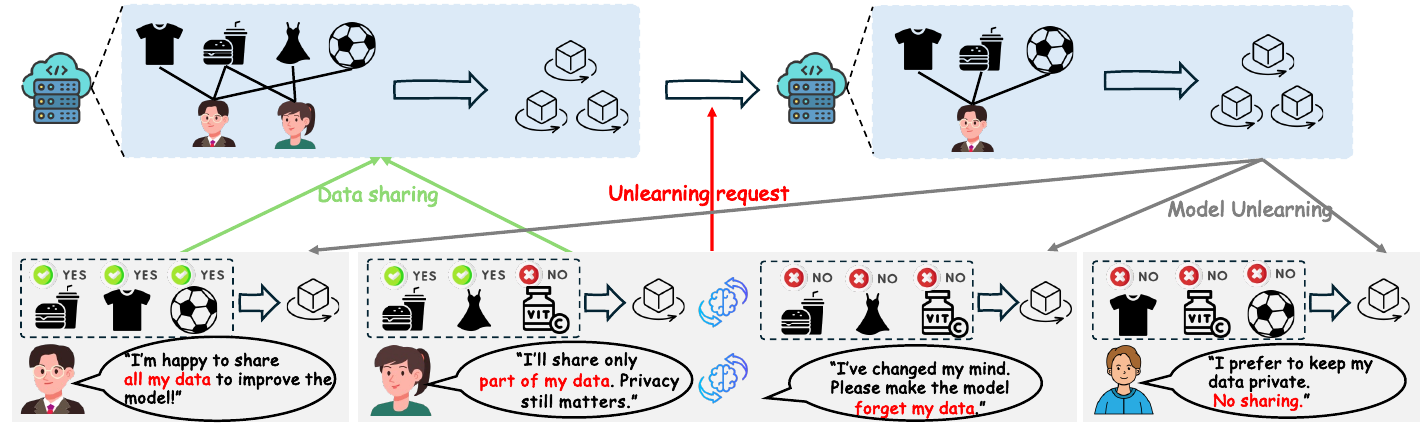}
  }

  \caption{Comparison of three federated recommendation architectures: (a) traditional model-sharing, (b) personalized data-sharing without unlearning, and (c) personalized data-sharing with unlearning support.}
  \label{fig:intro}
\end{figure*}

Inspired by the success of federated learning \cite{yang2019federated} in privacy protection, many recent studies have explored integrating federated learning into recommender systems, termed Federated Recommender Systems (FedRSs \cite{yang2020federated,sun2022survey,zhang2025personalized}). Specifically, as shown in Figure \ref{fig:intro_a}, FedRSs strictly keep users’ private data (e.g., user–item interaction data) on their local devices and deploy a local recommendation model for each user. In each training round, the server selects a group of users to participate in federated training. These selected users train their local models using local data and send only the model parameters to the server. The server then aggregates these parameters, instead of the raw data, into a global model using parameter aggregation algorithms (e.g., FedAvg \cite{mcmahan2017communication}) and redistributes the global model to all users. In this way, FedRSs can capture users’ preferences without sharing private user data. However, most existing FedRSs adopt a one-size-fits-all assumption regarding user privacy preferences, assuming that all users care strongly about data privacy. This assumption may not hold in practice, as it overlooks users who are less sensitive to privacy and are willing to share all or part of their data to obtain better recommendation services.

To address this limitation, several studies have explored personalized user data sharing mechanisms (as shown in Figure~\ref{fig:intro_b}) for federated recommender systems, which allow users to flexibly control whether and how their interaction data or profile attributes are shared with the server.
For example, CDCGNNFed~\cite{qu2024towards} enables users to choose to share all, partial, or no interaction data with the server, allowing the server to directly leverage richer collaborative information and thereby improve recommendation performance.
UC-FedRec~\cite{hu2024user} extends personalized data sharing to user attribute information, allowing users to selectively share profile attributes with the server.
PDC-FRS~\cite{yang2024pdc} further introduces a privacy-preserving data contribution mechanism that allows users to share interaction data under differential privacy guarantees.
Although these methods support personalized user data sharing, they all assume that users’ sharing preferences remain static once data are uploaded. In practice, however, user willingness to share data can be dynamic. As illustrated in Figure~\ref{fig:intro_c}, a user may regret sharing data after uploading it, or may share data by mistake. Therefore, it is necessary to support users to withdraw previously shared data and require the server to remove not only the data itself but also its influence on the trained model. This requirement is also aligned with the “Right to be Forgotten” in the GDPR.

One natural solution is to retrain federated recommender system models from scratch by removing the data that users request to be forgotten. However, although this solution can completely remove the influence of the data, it is very time-consuming and impractical in real-world federated recommender systems.
To address this limitation, recent work has begun to explore federated recommendation unlearning techniques, which aim to remove the influence of data without fully retraining the models by calibrating historical gradient information.
For example, FRU \cite{yuan2023federated} maintains important historical gradient information of model updates on each client. After receiving an unlearning request, all remaining clients reconstruct the unlearned models through gradient calibration. Similarly, CUFRU \cite{li2025cross} introduces a gradient transfer station, which leverages not only historical gradient information but also newly updated gradients from the remaining clients to improve gradient calibration.
However, existing federated recommendation unlearning methods require clients or the server to store historical gradient information, which introduces significant storage overhead and is especially challenging for resource-limited user devices.

In light of these challenges, this paper proposes a Federated Learn–Unlearn Framework for Recommender Systems with Personalized User Data Sharing, named FedShare.
Specifically, we introduce a fully personalized data-sharing mechanism that not only allows users to flexibly control whether their data are shared with the server, but also supports users in submitting unlearning requests to remove the influence of previously shared data from the trained model.
To achieve this goal, we introduce contrastive learning and contrastive unlearning to handle personalized data sharing and unlearning requests, respectively. When users choose to share data with the server, we construct a high-order user–item bipartite graph on the server side based on the shared user data. This graph can be regarded as an augmentation of the local first-order user–item subgraphs, which only contain each user and their interacted items. As a result, for each item, its representation is learned from two complementary views: a local view, obtained from local training via standard federated model aggregation techniques (e.g., FedAvg), and a global view, derived from the high-order user–item bipartite graph on the server. Contrastive learning is employed to align local and global item representations, thereby improving recommendation performance.

When users submit unlearning requests, the server first removes the data that need to be forgotten, resulting in an updated high-order user–item bipartite graph on the server side. However, it is non-trivial to directly apply existing federated recommendation unlearning techniques to remove the influence of these data, as they typically require storing substantial historical gradient information, which leads to significant storage overhead and poses serious challenges for both resource-limited user devices and the server. To alleviate this challenge, we propose a memory-efficient contrastive unlearning method that only requires the server to maintain few historical item embedding tables. Specifically, after data removal, the remaining users perform standard federated training to obtain updated global embedding tables, which are regarded as new local views. Meanwhile, on the server side, we construct a forgotten user–item bipartite graph based on the data to be forgotten and infer the corresponding item embeddings using the historical embedding tables, yielding forgotten views. In addition, by leveraging the updated high-order user–item bipartite graph constructed from the remaining data, the server derives new global views. The proposed contrastive unlearning strategy is then applied to these views, where new local views are aligned with new global views to preserve performance on the remaining data, while being selectively pushed away from forgotten views that capture the representations induced by the data to be forgotten, thereby ensuring that representations associated with the remaining data are not affected.

Overall, the contributions of this paper are summarized as follows:
\begin{itemize}
    \item To the best of our knowledge, this is the first work to explore fully personalized user data sharing mechanisms in federated recommender systems. It not only allows users to flexibly control the extent to which their data are shared, but also supports users in requesting the model to forget previously shared data.
    \item We propose a federated learn-unlearn framework for FedRSs, which introduces a dual contrastive learning technique. Compared with existing federated recommendation unlearning methods, the proposed framework significantly reduces storage overhead.
    \item We conduct extensive experiments on three benchmark datasets to evaluate the proposed method. The results show that FedShare achieves strong learning performance and efficient unlearning ability.
\end{itemize}

The remainder of this paper is organized as follows. Section \ref{sec:relatedwork} reviews related work and identifies the research gap. Section \ref{sec:preliminaries} introduces the preliminaries and formulates the problem. Section \ref{sec:method} presents the proposed method. Section \ref{sec:experi} shows the experimental results, and Section \ref{sec:con} concludes the paper.

\section{Related work}\label{sec:relatedwork}
In this section, we will first review the related work on federated recommender systems, including the recent studies that explore allowing users to freely control whether to share data with the server. Then, given that this work allows users to submit unlearning requests for data that needs to be forgotten, we will also discuss the related work on federated unlearning, especially studies that address unlearning in federated recommender systems.

\subsection{Federated recommender systems}
To address users’ privacy concerns in recommender systems, recent studies explore integrating federated learning into recommender systems, termed federated recommender systems \cite{ali2024hidattack,zhang2023comprehensive,gu2025gcpa,li2024fedcore,yin2025device,qu2025proxy,yuan2025robust}. Early work mainly focuses on matrix factorization–based federated recommender systems \cite{ammad2019federated,liu2022fairness,anelli2021federank,chai2020secure,lin2020fedrec,10.1145/3394486.3403176}, which learn user and item embeddings from local user–item interaction matrices. For example, FCF \cite{ammad2019federated} is one of the earliest works that integrates federated learning with matrix factorization, where user embeddings are learned on the client side, while a global item embedding model is maintained and updated on the server side. FedMF \cite{chai2020secure} further incorporates homomorphic encryption into the parameter sharing process to enhance privacy protection. FedFast \cite{10.1145/3394486.3403176} leverages clustering techniques to group users and selects clients from different groups in each communication round, thereby improving the convergence speed of federated training. To better capture the nonlinear relationships between users and items, subsequent studies explore deep neural network–based federated recommender systems \cite{zhang2023dual,10.1145/3397271.3401081,10.1016/j.knosys.2022.108441}. For example, FedNCF \cite{10.1016/j.knosys.2022.108441} adopts NeuMF \cite{he2017neuralcol} as the base recommendation model, which combines multilayer perceptrons and matrix factorization to more effectively learn latent user–item interactions. PFedRS \cite{zhang2023dual} proposes a personalized federated recommendation model by learning personalized local similarity functions for different users. In addition, some studies model user–item interactions as a bipartite graph and employ graph neural networks as the base recommendation model. For instance, FedGNN \cite{wu2022federated} introduces a trusted third-party server to construct high-order graphs, enabling GNNs to learn user and item embeddings in a privacy-preserving manner. SemiDFEGL \cite{qu2023semi} further applies clustering techniques to both users and items to construct high-order local subgraphs, thereby capturing higher-order structural information and improving recommendation performance.

However, although existing federated recommender systems can effectively protect user privacy while providing satisfactory recommendation performance, most of them adopt a one-size-fits-all privacy budget assumption, i.e., assuming that all users have equally strong privacy concerns. This assumption ignores users who are willing to share all or part of their data in exchange for better recommendation quality. In recent years, only a limited number of studies have explored federated recommender systems that support personalized user data sharing. For example, CDCGNNFed \cite{qu2024towards} and UC-FedRec \cite{hu2024user} are two pioneering works that allow users to freely share user–item interaction data and attribute data, respectively. Furthermore, PDC-FRS \cite{yang2024pdc} introduces a privacy-preserving mechanism that enables users to share data with differential privacy guarantees. Despite these advances, existing methods only support data sharing and overlook users’ dynamic privacy preferences, such as scenarios where users later regret their decisions and request the server to delete or even unlearn the influence of previously shared data. To fill this gap, this work aims to support fully personalized user data sharing, which not only allows flexible data sharing but also enables unlearning requests in federated recommender systems.

\subsection{Federated recommendation unlearning}
To remove the influence of data that needs to be forgotten from a trained model, the most straightforward solution is to retrain the model from scratch after removing the target data. However, this solution is clearly time-consuming and impractical in real-world applications. To address this issue, recent studies explore removing the influence of forgotten data without retraining the model from scratch, giving rise to machine unlearning \cite{tarun2023fast,wang2024machine,bourtoule2021machine,nguyen2025survey}, including unlearning methods for recommendation tasks \cite{chen2022recommendation,li2024survey}. Typical methods include fine-tuning–based methods \cite{chowdhury2024towards}, which update the model using the remaining data; gradient ascent–based methods \cite{hua2024federated}, which perform inverse learning on the data to be forgotten; and data synthesis–based methods \cite{wang2024mitigating}, which modify the labels of the forgotten data and then fine-tune the model.
However, these methods are generally not applicable to federated learning settings, as the data to be removed cannot be directly accessed due to privacy concerns. To address this challenge, several studies explore federated recommendation unlearning \cite{liu2024survey} by leveraging clients’ historical gradient information to reconstruct the model after unlearning requests are issued. For example, FRU \cite{yuan2023federated} maintains historical local gradients at each client and introduces an importance sampling strategy to selectively retain important gradients, thereby reducing storage costs and enabling model reconstruction via gradient calibration after unlearning requests. Similarly, CUFRU \cite{li2025cross} introduces a gradient transfer station that leverages not only historical gradients but also newly updated gradients from the remaining clients to improve gradient calibration. In addition, CFRU \cite{huynh2025certified} proposes a gradient sampling strategy to further reduce storage costs. Furthermore, Aegis \cite{wu2025aegis} adopts a post-training method to perform attribute unlearning in federated recommender systems.

Despite their effectiveness, existing federated recommendation unlearning methods either require clients or the server to store a large amount of historical gradient information, leading to significant storage overhead. This issue is particularly challenging for resource-constrained client devices. To address this limitation, we explore memory-efficient federated recommendation unlearning, which does not require storing historical gradients on clients and only maintains a small number of historical parameters on the server.

\section{Preliminaries}\label{sec:preliminaries}

In this section, we first introduce the typical pipeline of federated recommender systems. Then, we formally define the personalized user data sharing problem in federated recommender systems. The commonly used symbols are summarized in Table~\ref{tab:symbols}.

\begin{table}[t]
\centering
\caption{Summary of notations.}
\label{tab:symbols}
\begin{tabular}{ll}
\toprule
\textbf{Notation} & \textbf{Description} \\
\midrule
$\mathcal{U}$ & Set of users (clients) \\
$\mathcal{I}$ & Set of items \\
$\mathcal{I}_u$ & Items interacted by user $u$ \\
$\mathcal{I}_u^{\mathrm{local}}$ & Interactions kept local by user $u$ \\
$\mathcal{I}_u^{\mathrm{share}}$ & Interactions shared with the server by user $u$ \\
$\mathcal{I}_u^{\mathrm{unlearn}}$ & Shared interactions requested to be unlearned \\
$\mathcal{D}_u$ & Local interaction dataset of user $u$ \\
$\mathcal{D}_s$ & Server-side dataset constructed from shared interactions \\
\midrule
$d$ & Embedding dimension \\
$\mathbf{u}_u$ & User embedding of user $u$ \\
$\mathbf{v}_i$ & Item embedding of item $i$ \\
$\mathbf{V}_u$ & Local item embedding table on client $u$ \\
$\mathbf{V}_s$ & Global item embedding table on the server \\
$\boldsymbol{\theta}_u$ & Local non-embedding model parameters \\
$\boldsymbol{\theta}_s$ & Global non-embedding model parameters \\
$\Theta_u$ & Local model parameters of user $u$ \\
$\Theta_s$ & Global model parameters on the server \\
$\tilde{\Theta}_s$ & Global model parameters after unlearning \\
$\Theta_s^{\star}$ & Ideal retrained global model parameters \\
\midrule
$g(\cdot,\cdot)$ & User--item scoring (similarity) function \\
$\hat{y}_{u,i}$ & Predicted preference score of user $u$ for item $i$ \\
$l_u(\cdot)$ & Local loss function of user $u$ \\
$\mathrm{Agg}(\cdot)$ & Aggregation function in federated training \\
$\mathcal{F}_{\mathrm{un}}(\cdot)$ & Unlearning function on the global model \\
\bottomrule
\end{tabular}
\end{table}

\subsection{Federated Recommender Systems}
Let $\mathcal{U}$ be the set of users/clients\footnote{We interchangeably use the terms ``user'' and ``client''
based on context, and assume each user corresponds to a single client device.} and $\mathcal{I}$ the set of items. 
In the context of federated recommender systems, each user $u \in \mathcal{U}$ holds a 
private local interaction dataset $\mathcal{D}_{u}$ defined as
\begin{equation}
\mathcal{D}_u = \{\, i \in \mathcal{I}_{u} \,\},
\end{equation}
where $\mathcal{I}_{u} \subseteq \mathcal{I}$ denotes the set of items that user $u$ has interacted with. Typically, $\mathcal{D}_u$ is strictly retained on the user side due to privacy concerns.

In addition, each client $u \in \mathcal{U}$ maintains a local recommender model
$f_u(\cdot)$.
Because recommender systems are typically \emph{embedding-based} and represent users and items as $d$-dimensional vectors, the local model on client $u$ generally consists of three types of parameters:
(i) a user embedding $\mathbf{u}_u \in \mathbb{R}^{d}$, which is usually kept locally due to privacy concerns;
(ii) a local item embedding table $\mathbf{V}_u \in \mathbb{R}^{|\mathcal{I}| \times d}$;
and (iii) model parameters $\boldsymbol{\theta}_u$ that are independent of embeddings
(e.g., the weights of neural network layers).
For convenience, we denote the complete set of local model parameters on client $u$ as
$\Theta_u = \{\mathbf{u}_u, \mathbf{V}_u, \boldsymbol{\theta}_u\}$.
Correspondingly, the server maintains a global model parameterized by
$\Theta_s = \{\mathbf{V}_s, \boldsymbol{\theta}_s\}$, which includes the global item
embedding table $\mathbf{V}_s \in \mathbb{R}^{|\mathcal{I}| \times d}$ and global model parameters $\boldsymbol{\theta}_s$.

A typical federated training process proceeds in rounds.
At the beginning of round $t$, the server selects a subset of clients
$\mathcal{U}^{(t)} \subseteq \mathcal{U}$ and broadcasts the current global parameters
$\Theta_s^{(t-1)}=\{\mathbf{V}_s^{(t-1)},\boldsymbol{\theta}_s^{(t-1)}\}$ to all selected clients:
\begin{equation}
\mathbf{V}_u^{(t)} \leftarrow \mathbf{V}_s^{(t-1)}, \qquad
\boldsymbol{\theta}_u^{(t)} \leftarrow \boldsymbol{\theta}_s^{(t-1)}, \qquad
\forall u \in \mathcal{U}^{(t)}.
\end{equation}
Note that the user embedding $\mathbf{u}_u$ is kept locally and is not uploaded to the server.

Each selected client $u$ then performs local training using its private data $\mathcal{D}_u$.
Given a user embedding $\mathbf{u}_u$ and an item embedding $\mathbf{v}_i$,
the predicted preference score between user $u$ and item $i$ is computed as
$\hat{y}_{u,i} = g(\mathbf{u}_u, \mathbf{v}_i)$, where $g(\cdot,\cdot)$ denotes a
similarity or scoring function.
Let $l_u(\Theta_u;\mathcal{D}_u)$ denote the local loss on client $u$.
Client $u$ can update its local parameters as follows:
\begin{equation}
\Theta_u^{(t)} \leftarrow \Theta_u^{(t)} - \eta \nabla_{\Theta_u}l_u\!\left(\Theta_u^{(t)};\mathcal{D}_u\right),
\end{equation}
where $\eta$ is the learning rate.
After local training, each client sends only the updated item embedding table
$\mathbf{V}_u^{(t)}$ and model parameters $\boldsymbol{\theta}_u^{(t)}$ to the server, without
sharing the raw interaction data.

The server aggregates the received updates to obtain the new global parameters using
an aggregation function $\mathrm{Agg}(\cdot)$:
\begin{equation}
\Theta_s^{(t)} = 
\mathrm{Agg}\!\left( \{\Theta_u^{(t)} \mid u \in \mathcal{U}^{(t)}\} \right).
\end{equation}

Overall, the goal of federated recommender systems is to minimize the sum of local objectives:
\begin{equation}
\min_{\{\Theta_u\}}
\sum_{u\in\mathcal{U}} l_u\!\left(\Theta_u;\mathcal{D}_u\right).
\end{equation}

\subsection{Federated recommender systems with personalized user data sharing}

The above traditional federated recommender systems assume that all user interaction data remain strictly local and are never shared with the server. However, users may have different privacy preferences in practice, and some users may be willing to share all or part of their interaction data in exchange for improved recommendation performance.
This motivates the study of federated recommender systems with \emph{personalized user data sharing}, where users can flexibly control which data to keep local, which data to share with the server, and which shared data should later be removed, together with removing their influence on the trained model.

Formally, for each user $u \in \mathcal{U}$, to model personalized data-sharing
preferences, the local interaction dataset is defined as
\begin{equation}
    \mathcal{D}_u =
\big(
\mathcal{I}_u^{\mathrm{local}},
\mathcal{I}_u^{\mathrm{share}},
\mathcal{I}_u^{\mathrm{unlearn}}
\big),
\end{equation}
where $\mathcal{I}_u^{\mathrm{local}}$ denotes interactions that are kept strictly local and
never shared with the server,
$\mathcal{I}_u^{\mathrm{share}}$ denotes interactions that users are willing to share with the
server, and
$\mathcal{I}_u^{\mathrm{unlearn}} \subseteq \mathcal{I}_u^{\mathrm{share}}$ denotes shared
interactions whose influence the user later requests to be removed. 
Correspondingly, the server can constructs a server-side dataset based on all shared
interactions, defined as
\begin{equation}
    \mathcal{D}_s = \bigcup_{u \in \mathcal{U}} \mathcal{I}_u^{\mathrm{share}}
\end{equation}

Under this setting, the system still aims to generate personalized top-$k$ recommendation lists
for users based on the learned user and item embeddings, while learning from the
available interaction data
$\mathcal{I}_u^{\mathrm{local}} \cup \mathcal{I}_u^{\mathrm{share}}$.

When unlearning requests are issued, the server-side dataset becomes
$\mathcal{D}_s \setminus \bigcup_{u \in \mathcal{U}} \mathcal{I}_u^{\mathrm{unlearn}}$.
Accordingly, the trained model needs to be updated to remove the influence of
$\mathcal{I}_u^{\mathrm{unlearn}}$.

Formally, we define an unlearning function $\mathcal{F}_{\mathrm{un}}(\cdot)$ that operates
on the current global model parameters.
After unlearning, the updated global model is given by
\begin{equation}
\tilde{\Theta}_s
=
\mathcal{F}_{\mathrm{un}}\!\left(
\Theta_s,\,
\{\mathcal{I}_u^{\mathrm{unlearn}}\}_{u \in \mathcal{U}}
\right),
\end{equation}
where $\tilde{\Theta}_s$ denotes the global model parameters after unlearning.

The goal of unlearning is to ensure that the updated global model $\tilde{\Theta}_s$ is
consistent with an ideal retrained global model $\Theta_s^{\star}$, which is obtained by
retraining the federated recommender system from scratch using the remaining interaction
data
$\mathcal{I}_u^{\mathrm{local}} \cup
(\mathcal{I}_u^{\mathrm{share}} \setminus \mathcal{I}_u^{\mathrm{unlearn}})$ for all users
$u \in \mathcal{U}$.
That is,
\begin{equation}
\tilde{\Theta}_s \;\approx\; \Theta_s^{\star}.
\end{equation}

\section{FedShare}\label{sec:method}
\begin{figure*}
    \centering
    \includegraphics[width=0.8\linewidth]{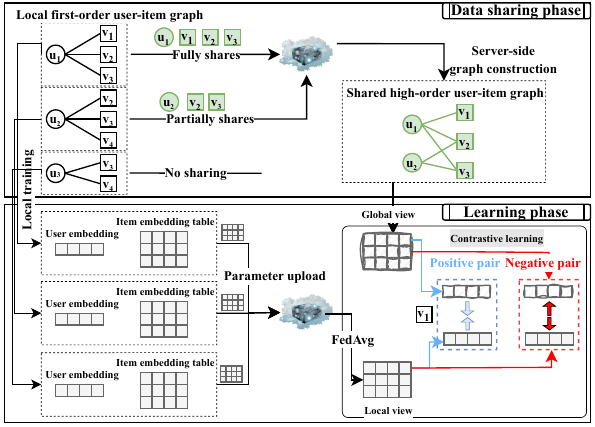}
    \caption{The learning phase of FedShare, where users share interaction data according to personalized preferences to construct a server-side high-order user–item graph, and contrastive learning is applied to align local and global item representations.}
    \label{fig:learningphase}
\end{figure*}

\begin{figure*}
    \centering
    \includegraphics[width=0.8\linewidth]{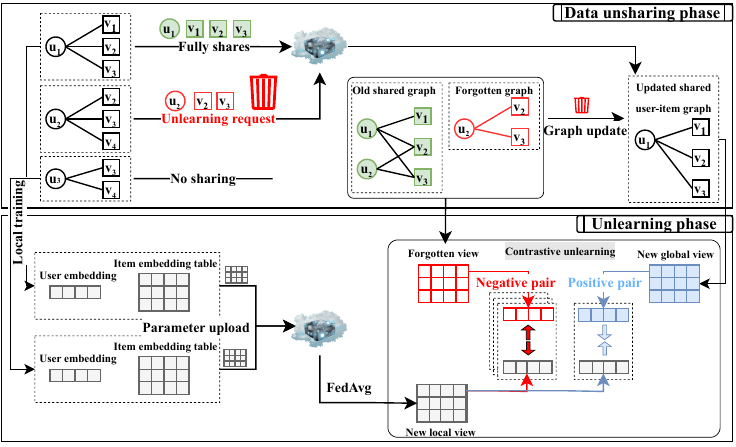}
    \caption{The unlearning phase of FedShare, where unlearning requests update the shared user–item graph and contrastive unlearning removes the influence of forgotten data.}
    \label{fig:unlearnining phase}
\end{figure*}

In this section, we delve into the details of the proposed \textbf{FedShare} framework.
As illustrated in Fig.~\ref{fig:learningphase} and Fig.~\ref{fig:unlearnining phase}, the overall
framework consists of two main phases: a \emph{learning phase} based on personalized
user data sharing, and an \emph{unlearning phase} that removes the influence of data
users request to forget.
\textbf{(1) Learning phase.}
The learning phase starts with personalized data sharing, where users can freely choose
to share all, part, or none of their interaction data with the server.
Based on the data shared by users, the server constructs a server-side high-order
user-item bipartite graph.
To better exploit the rich relational information contained in this graph, the server
derives a \emph{global view} of item embeddings from the high-order graph, while
clients obtain a \emph{local view} of item embeddings through local training and
federated aggregation.
FedShare then employs contrastive learning to refine item embeddings by aligning the
local and global views, thereby improving the quality of learned representations.
\textbf{(2) Unlearning phase.}
When users submit unlearning requests, the server first removes the corresponding
interactions from the server-side high-order user--item graph.
After graph update, we introduce a contrastive unlearning technique to efficiently
remove the influence of the unlearned data.
Specifically, new local views obtained from federated training on the remaining users
are encouraged to align with new global views derived from the updated server-side graph,
while being pushed away from \emph{forgotten views} that are constructed based on the
removed data. In this way, FedShare selectively removes the
impact of unlearned data while preserving model performance on the remaining data.

\subsection{Data Sharing}

Traditional federated recommender systems adopt a one-size-fits-all privacy setting,
where all users are required to keep their interaction data strictly local.
While this design provides strong privacy protection, it overlooks users who are willing
to share all or part of their interaction data with the server in exchange for improved
recommendation performance, leading to unnecessary performance degradation.
From a graph perspective, this limitation arises because each client is limited to a user-centric first-order user–item graph.
Specifically, for each user $u$, the local interaction data can be represented as a
first-order bipartite graph
\begin{equation}
\mathcal{G}_u = (\{u\} \cup \mathcal{I}_u,\; \mathcal{E}_u),
\end{equation}
where $\mathcal{E}_u = \{(u,i)\mid i \in \mathcal{I}_u\}$ denotes interactions between user
$u$ and the items they have interacted with.
As illustrated in Fig.~\ref{fig:learningphase}, such strictly local graphs prevent the server
from directly exploiting high-order collaborative signals across users \cite{qu2023semi}, which are
commonly leveraged in centralized recommender systems \cite{he2020lightgcn}.

To address this limitation, we introduce a \emph{personalized user data sharing} mechanism
for federated recommender systems, which allows each user to flexibly control how much
interaction data is shared with the server.
Under personalized data sharing, we define
$\mathcal{I}_u^{\mathrm{share}} \subseteq \mathcal{I}_u$ as the subset of interactions that
user $u$ chooses to share with the server.
This definition naturally covers different user preferences:
when $\mathcal{I}_u^{\mathrm{share}} = \mathcal{I}_u$, user $u$ fully shares all interaction
data;
when $\mathcal{I}_u^{\mathrm{share}}$ is a non-empty proper subset of $\mathcal{I}_u$, user
$u$ partially shares interaction data;
and when $\mathcal{I}_u^{\mathrm{share}} = \varnothing$, no data are shared with the server,
which reduces to the conventional federated recommender setting.

Based on the interactions shared by users, the server constructs a server-side dataset$
 \mathcal{D}_s = \bigcup_{u \in \mathcal{U}} \mathcal{I}_u^{\mathrm{share}},
$
which can be viewed as a shared high-order user--item bipartite graph
\begin{equation}
\mathcal{G}_s = (\mathcal{U}_{\mathrm{share}} \cup \mathcal{D}_s,\; \mathcal{E}_s),
\end{equation}
where $\mathcal{U}_{\mathrm{share}} = \{u \mid \mathcal{I}_u^{\mathrm{share}} \neq \varnothing\}$
denotes the set of users who share data, and
$\mathcal{E}_s = \{(u,i)\mid i \in \mathcal{I}_u^{\mathrm{share}}\}$.
This server-side graph allows the server to directly leverage collaborative information across users that is unavailable in traditional federated recommender systems.

\subsection{Federated learning phase}
After the server constructs the high-order user--item graph $\mathcal{G}_s$ based on the
shared interaction data, FedShare proceeds with federated training as follows.
The overall training procedure is summarized in Algorithm~\ref{alg:fedshare_learning}.

\begin{algorithm}[t]
\caption{Learning Phase of FedShare}
\label{alg:fedshare_learning}
\begin{algorithmic}[1]
\Require High-order server-side graph $\mathcal{G}_s$; initial global parameters $\Theta_s^{(0)}=\{\mathbf{V}_s^{(0)},\boldsymbol{\theta}_s^{(0)}\}$; total rounds $T$
\For{$t=1,2,\ldots,T$}
    \State Server samples a client subset $\mathcal{U}^{(t)} \subseteq \mathcal{U}$
    \State Server broadcasts $\{\mathbf{V}_s^{(t-1)},\boldsymbol{\theta}_s^{(t-1)}\}$ to all $u\in\mathcal{U}^{(t)}$
    \ForAll{$u\in\mathcal{U}^{(t)}$ \textbf{in parallel}}
        \State Initialize $\mathbf{V}_u^{(t)}\leftarrow \mathbf{V}_s^{(t-1)}$, $\boldsymbol{\theta}_u^{(t)}\leftarrow \boldsymbol{\theta}_s^{(t-1)}$
        \State Infer local embedding using Eq.~\eqref{eq:lgc_general} and Eq.~\eqref{eq:layer_comb}
        \State Optimize local model with BPR in Eq.~\eqref{eq:bpr}
        \State Upload $\mathbf{V}_u^{(t)}$ to the server
    \EndFor

    \State Aggregate client updates via Eq.~\eqref{eq:fedavg}

    \State Infer item embeddings on $\mathcal{G}_s$ via Eq.~\eqref{eq:lgc_general} and Eq.~\eqref{eq:layer_comb}
    \State Update $\mathbf{V}_s^{(t)}$ via Eq.~\eqref{eq:serverlearning}
\EndFor
\end{algorithmic}
\end{algorithm}

\subsubsection{Client Selection}
At communication round $t$, the server randomly selects a subset of clients,
denoted as $\mathcal{U}^{(t)} \subseteq \mathcal{U}$, to participate in the current round
of federated training.
The server then distributes the global model parameters obtained from the previous round
to the selected clients as initialization for local training.
Specifically, the global item embedding table and model parameters are assigned as
\begin{equation}
\mathbf{V}_u^{(t)} \leftarrow \mathbf{V}_s^{(t-1)}, \qquad
\boldsymbol{\theta}_u^{(t)} \leftarrow \boldsymbol{\theta}_s^{(t-1)},
\quad \forall u \in \mathcal{U}^{(t)}.
\end{equation}
These parameters serve as the initialization for subsequent local model training on each
selected client.

\subsubsection{Local Training}\label{sec:localtraining}

After receiving the global model parameters, each selected client
$u \in \mathcal{U}^{(t)}$ performs local training based on its private interaction data
$\mathcal{D}_u$.
It is worth noting that local training is performed only on the remaining interactions after data sharing, which effectively reduces the training overhead for users who choose to share part of their data with the server.

Inspired by the success of graph-based recommender systems \cite{wu2022graph}, we adopt
LGC~\cite{he2020lightgcn} as the backbone model for local training.
LGC models the recommender system as a user--item bipartite graph, where users and
items are treated as distinct nodes and embeddings are updated by propagating information
from neighboring nodes.
Formally, for client $u$, the local interaction data can be represented as a user-centric
first-order user--item bipartite graph consisting of the user $u$ and the items
$\mathcal{I}_u$ they have interacted with.
Following LGC, the embedding propagation at layer $l+1$ is defined as
\begin{equation}
\begin{aligned}
\mathbf{u}_u^{(l+1)}
&=
\sum_{i \in \mathcal{N}(u)}
\frac{1}{\sqrt{|\mathcal{N}(u)|}\sqrt{|\mathcal{N}(i)|}}
\, \mathbf{v}_i^{(l)}, \\
\mathbf{v}_i^{(l+1)}
&=
\sum_{u' \in \mathcal{N}(i)}
\frac{1}{\sqrt{|\mathcal{N}(i)|}\sqrt{|\mathcal{N}(u')|}}
\, \mathbf{u}_{u'}^{(l)} .
\end{aligned}
\label{eq:lgc_general}
\end{equation}
where $\mathcal{N}(u)$ and $\mathcal{N}(i)$ denote the sets of neighboring items and users,
respectively, and $\mathbf{u}_u^{(l)}$ and $\mathbf{v}_i^{(l)}$ are the user and item
embeddings at layer $l$.

After $L$ propagation layers, LGC adopts a layer combination strategy to obtain the
final user and item representations.
Specifically, the final embeddings are computed as
\begin{equation}
\mathbf{u}_u
=
\sum_{l=0}^{L} \alpha_l \mathbf{u}_u^{(l)},
\qquad
\mathbf{v}_i
=
\sum_{l=0}^{L} \alpha_l \mathbf{v}_i^{(l)},
\label{eq:layer_comb}
\end{equation}
where $\alpha_l$ denotes the importance weight of the $l$-th propagation layer. In the federated setting, each client only maintains a user-centric first-order ego graph.
As a result, for client $u$, we have $\mathcal{N}(u)=\mathcal{I}_u$, and for each item
$i \in \mathcal{I}_u$, its neighborhood contains only user $u$, i.e.,
$\mathcal{N}(i)=\{u\}$ and $|\mathcal{N}(i)|=1$.
Under this setting, higher-order message passing is not available on the client side.
Therefore, we employ a single propagation layer ($L=1$) for local embedding inference.

For local model optimization, we adopt the widely used Bayesian Personalized Ranking
(BPR) loss~\cite{rendle2012bpr}.
For each user $u$, the local training objective is defined as
\begin{equation}
\mathcal{L}_{\mathrm{BPR}}^{(u)}
=
- \sum_{i \in \mathcal{I}_u}
\sum_{j \notin \mathcal{I}_u}
\ln
\sigma\!\left(
\cos(\mathbf{u}_u,\mathbf{v}_i)
-
\cos(\mathbf{u}_u,\mathbf{v}_j)
\right)
+
\lambda \|\Theta_u\|_2^2,
\label{eq:bpr}
\end{equation}
where $\sigma(\cdot)$ denotes the sigmoid function and $\lambda$ controls the strength of
$\ell_2$ regularization.

\subsubsection{Global Model Update}

After local training, each participating client uploads its updated model parameters to
the server.
Specifically, clients only transmit the updated item embedding tables
$\mathbf{V}_u^{(t)}$, while user embeddings $\mathbf{u}_u$ are kept locally due to privacy
concerns.
Based on the received updates from clients in round $t$, the server performs model
aggregation to obtain a local-view item embedding table.
In this work, we adopt the widely used FedAvg \cite{mcmahan2017communication} as the
aggregation function.
Specifically, the server updates the global item embedding table as
\begin{equation}
\mathbf{V}_s^{(t)}
=
\sum_{u \in \mathcal{U}^{(t)}}
\frac{|\mathcal{D}_u|}{\sum_{v \in \mathcal{U}^{(t)}} |\mathcal{D}_v|}
\, \mathbf{V}_u^{(t)},
\label{eq:fedavg}
\end{equation}
where $|\mathcal{D}_u|$ denotes the number of local interactions held by client $u$.
This aggregation yields a \emph{local-view} item embedding table that summarizes knowledge
learned from local training.

While the aggregated item embeddings capture user--item interactions from local views,
they cannot fully exploit the high-order collaborative signals embedded in the
server-side user--item graph $G_s$ constructed from shared data.
To better leverage such high-order structural information, we further refine the item
embeddings using contrastive learning on $G_s$.
Specifically, for each item $i$ in the server-side graph $G_s$, we construct two
complementary views.
On the one hand, its \emph{local view} representation $\mathbf{v}_i^{\mathrm{local}}$ is
directly retrieved from the aggregated item embedding table $\mathbf{V}_s^{(t)}$.
On the other hand, its \emph{global view} representation $\mathbf{v}_i^{\mathrm{global}}$
is obtained by performing LGC-based embedding propagation on $G_s$, following
Eq.~(\ref{eq:lgc_general}) and the layer combination in Eq.~(\ref{eq:layer_comb}).

Based on these two views, we treat $(\mathbf{v}_i^{\mathrm{local}}, \mathbf{v}_i^{\mathrm{global}})$
as a positive pair for item $i$, while embeddings of other items within the same batch are used as negative samples.
Following SimCLR~\cite{chen2020simple}, we adopt the InfoNCE loss~\cite{gutmann2010noise}
for contrastive learning:
\begin{equation}
\mathcal{L}_{\mathrm{CL}}
=
- \sum_{i \in \mathcal{I}_s}
\log
\frac{
\exp\!\left( \cos(\mathbf{v}_i^{\mathrm{local}}, \mathbf{v}_i^{\mathrm{global}}) / \tau \right)
}{
\sum_{j \in \mathcal{I}_s}
\exp\!\left( \cos(\mathbf{v}_i^{\mathrm{local}}, \mathbf{v}_j^{\mathrm{global}}) / \tau \right)
},
\end{equation}
where $\mathcal{I}_s$ denotes the set of items in $G_s$ and $\tau$ is the
temperature parameter.
In addition to contrastive learning, we further optimize the server-side model using the BPR loss on $G_s$ to directly capture high-order collaborative signals.
Overall, the final server-side loss function is defined as
\begin{equation}
\mathcal{L}
=
\mathcal{L}_{\mathrm{BPR}}
+
\lambda_1 \mathcal{L}_{\mathrm{CL}}
+
\lambda \|\mathbf{V}_s\|_2^2,
\label{eq:serverlearning}
\end{equation}
where $\lambda_1$ and $\lambda$ control the strengths of contrastive learning and
regularization, respectively.

\subsection{Data Unsharing}
In practical systems, users’ privacy preferences may change over time.
A user who previously shared interaction data with the server may later request that part or all of the shared data be removed, which we refer to as \emph{forgotten data}.
We term this operation \emph{data unsharing}, which requires not only removing the corresponding data from the server, but also removing their influence on the trained model.

Formally, for each user $u$, we define $\mathcal{I}_u^{\mathrm{unlearn}} \subseteq
\mathcal{I}_u^{\mathrm{share}}$ as the set of shared interactions that the user requests to
unshare.
Once unsharing requests are issued, the server removes these interactions from the
server-side dataset.
Accordingly, the server-side dataset and graph are updated as
\begin{equation}
\mathcal{D}_s \leftarrow
\mathcal{D}_s \setminus \bigcup_{u \in \mathcal{U}} \mathcal{I}_u^{\mathrm{unlearn}},
\end{equation}
and
\begin{equation}
\mathcal{G}_s \leftarrow
\mathcal{G}_s \setminus
{(u,i) \mid i \in \mathcal{I}_u^{\mathrm{unlearn}},, u \in \mathcal{U}},
\end{equation}
respectively.

It is important to note that data unsharing only affects interactions that were previously
shared with the server.
All interactions in $\mathcal{I}_u^{\mathrm{local}}$ remain strictly on the client side and
are never exposed to the server throughout the training or unlearning process.

\subsection{Federated Unlearning Phase}

After data unsharing, the server updates the shared dataset and graph by removing the
requested interactions, i.e.,
$\mathcal{D}_s \leftarrow \mathcal{D}_s \setminus \bigcup_{u \in \mathcal{U}} \mathcal{I}_u^{\mathrm{unlearn}}$
and $\mathcal{G}_s \leftarrow \mathcal{G}_s \setminus \{(u,i)\mid i\in \mathcal{I}_u^{\mathrm{unlearn}}\}$.
However, simply removing edges from $\mathcal{G}_s$ does not remove their influence
from the trained model, since the current parameters already encode historical updates.
A straightforward solution is full retraining from scratch, but it is time-consuming and
impractical.
Therefore, the federated unlearning phase aims to efficient update the global model to
remove the influence of $\mathcal{I}_u^{\mathrm{unlearn}}$ while preserving performance on
the remaining data.

Formally, given the current global parameters $\Theta_s$ and unlearning requests
$\{\mathcal{I}_u^{\mathrm{unlearn}}\}_{u\in\mathcal{U}}$, the unlearning procedure produces
updated global parameters
\begin{equation}
\tilde{\Theta}_s
=
\mathcal{F}_{\mathrm{un}}\!\left(
\Theta_s,\,
\{\mathcal{I}_u^{\mathrm{unlearn}}\}_{u \in \mathcal{U}}
\right),
\label{eq:unlearning_map}
\end{equation}
where the goal is to make $\tilde{\Theta}_s$ consistent with an ideal retrained model
$\Theta_s^{\star}$ learned on the remaining data, i.e.,
\begin{equation}
\tilde{\Theta}_s \approx \Theta_s^{\star}.
\label{eq:unlearning_goal}
\end{equation}
In the following, we describe how FedShare realizes $\mathcal{F}_{\mathrm{un}}(\cdot)$ via
a contrastive unlearning strategy with three following steps.

\subsubsection{New local View from Remaining-Data Federated Training}

A key requirement of unlearning is to avoid degrading recommendation performance on the
remaining data.
To this end, we first perform standard federated training using only the remaining
client-side interactions.
This step produces updated item embeddings that capture the data distribution after
unsharing, serving as a \emph{new local view}.
Specifically, this step follows the same federated local training pipeline described in
Section~\ref{sec:localtraining}.
At each communication round, the server randomly selects a subset of clients to
participate in local training.
Each selected client performs local training on its remaining interactions and uploads
the updated item embedding table to the server.
The server then aggregates the received updates using FedAvg, as defined in
Eq.~\eqref{eq:fedavg}, to obtain the new local-view item embedding table.

\subsubsection{Forgotten Views from Historical Embeddings and Forgotten Graph}

Directly pushing item embeddings away from all historical representations may severely
damage the knowledge learned from remaining data.
Therefore, effective unlearning should be \emph{selective}, i.e., removing the influence
of unshared interactions while preserving representations supported by the remaining
data.
To this end, we explicitly construct \emph{forgotten views} that capture the embedding
components induced by the unshared interactions $\mathcal{I}_u^{\mathrm{unlearn}}$.

A straightforward solution would be to store and manipulate historical gradients or
optimization trajectories associated with the forgotten data.
However, such designs typically incur substantial storage overhead, requiring either
clients (e.g., FRU~\cite{yuan2023federated}) or the server (e.g., CUFRU~\cite{li2025cross})
to maintain a large amount of historical gradient information.
In contrast, FedShare adopts a lightweight alternative that only stores a small number of
historical global item embedding tables on the server, which is sufficient to recover the
representation space induced by the forgotten interactions.

Formally, the server maintains a set of historical global item embedding tables from
previous training rounds, denoted as
$\{\mathbf{V}_s^{(t-m)}\}_{m=1}^{M}$, where $M$ is a small constant.
Based on the unshared interactions, we construct a \emph{forgotten graph}
\begin{equation}
\mathcal{G}_{\mathrm{f}} =
(\mathcal{U}_{\mathrm{f}} \cup \mathcal{I}_{\mathrm{f}},\; \mathcal{E}_{\mathrm{f}}),
\qquad
\mathcal{E}_{\mathrm{f}}=\{(u,i)\mid i \in \mathcal{I}_u^{\mathrm{unlearn}}\},
\label{eq:forgotten_graph}
\end{equation}
where $\mathcal{I}_{\mathrm{f}} = \bigcup_{u\in\mathcal{U}}\mathcal{I}_u^{\mathrm{unlearn}}$
denotes the set of items involved in the unshared interactions.

We then perform LGC-based embedding inference on $\mathcal{G}_{\mathrm{f}}$ using each
historical embedding table $\mathbf{V}_s^{(t-m)}$ as initialization.
Specifically, by applying Eq.~\eqref{eq:lgc_general} and Eq.~\eqref{eq:layer_comb}, we obtain
a corresponding forgotten-view embedding table
$\mathbf{V}_{\mathrm{f}}^{(t-m)}$ for each historical snapshot.
As a result, each item $i \in \mathcal{I}_{\mathrm{f}}$ is associated with a set of
forgotten views
\begin{equation}
\mathcal{V}_{i,\mathrm{f}}^{(t)}
=
\left\{
\mathbf{v}_{i,\mathrm{f}}^{(t-m)}
\;\middle|\;
m = 1,\ldots,M
\right\},
\label{eq:forgotten_view_set}
\end{equation}
which characterize how the forgotten interactions shape the item representations.
These forgotten views are later used as negative samples in contrastive unlearning to
remove the influence of the unshared data, while preserving representations supported by
the remaining data.

\subsubsection{New Global Views from the Updated Shared Graph}

After removing unshared interactions, the updated shared graph $\mathcal{G}_s$ encodes
high-order collaborative signals from the remaining shared data.
To use these signals during unlearning, we infer \emph{new global views} on the updated
$\mathcal{G}_s$.

Specifically, we perform LGC propagation on the updated $\mathcal{G}_s$ using
Eq.~\eqref{eq:lgc_general} and Eq.~\eqref{eq:layer_comb}, with the new local-view embedding
table $\mathbf{V}_{s,\mathrm{loc}}^{(t)}$ as initialization, yielding a new global-view item
embedding table $\mathbf{V}_{s,\mathrm{glob}}^{(t)}$.
We denote the global-view embedding of item $i$ as $\mathbf{v}_{i,\mathrm{glob}}^{(t)}$.
\label{eq:new_global_view}

\subsubsection{Contrastive Unlearning}

After data unsharing, each item can be associated with up to three different
representations that capture its semantics from complementary perspectives.
Specifically, for item $i$, we consider:
(i) a \emph{new local view} $\mathbf{v}_{i,\mathrm{loc}}^{(t)}$, obtained from federated
training on the remaining client-side interactions;
(ii) a \emph{new global view} $\mathbf{v}_{i,\mathrm{glob}}^{(t)}$, inferred on the updated
server-side graph $\mathcal{G}_s$; and
(iii) a set of \emph{forgotten views} $\mathcal{V}_{i,\mathrm{f}}^{(t)}$, inferred from
historical embedding tables on the forgotten graph $\mathcal{G}_{\mathrm{f}}$.

The goal of unlearning is twofold:
on the one hand, item representations should remain consistent with the data retained
after unsharing;
on the other hand, representations induced by the unshared interactions should be
effectively removed.
To achieve this, we design a contrastive unlearning objective that simultaneously
\emph{aligns} the new local and new global views, while \emph{separating} the new local
view from the forgotten views.

Formally, for each item $i$, we treat the pair
$(\mathbf{v}_{i,\mathrm{loc}}^{(t)}, \mathbf{v}_{i,\mathrm{glob}}^{(t)})$
as a positive pair.
For items involved in unsharing requests, i.e., $i \in \mathcal{I}_{\mathrm{f}}$,
the corresponding forgotten views
$\mathcal{V}_{i,\mathrm{f}}^{(t)}$ are used as negative samples.
For items not appearing in forgotten interactions, no forgotten negatives are applied.

Following the InfoNCE formulation, the contrastive unlearning loss is defined as
\begin{equation}
\mathcal{L}_{\mathrm{CU}}
=
- \sum_{i \in \mathcal{I}_s}
\log
\frac{
\exp\!\left(
\cos(\mathbf{v}_{i,\mathrm{loc}}^{(t)}, \mathbf{v}_{i,\mathrm{glob}}^{(t)}) / \tau
\right)
}{
\sum\limits_{\mathbf{v} \in \mathcal{V}_{i,\mathrm{f}}^{(t)}}
\exp\!\left(
\cos(\mathbf{v}_{i,\mathrm{loc}}^{(t)}, \mathbf{v}) / \tau
\right)
},
\label{eq:contrastive_unlearning}
\end{equation}
where $\tau$ is the temperature parameter.

By minimizing $\mathcal{L}_{\mathrm{CU}}$, the model preserves item representations that
are supported by the remaining data through positive alignment, while reducing the
influence of representations associated with the unshared interactions via item-specific
forgotten negatives. The overall federated unlearning procedure is summarized in Algorithm \ref{alg:fedshare_unlearning}.

\begin{algorithm}[t]
\caption{Federated Unlearning Phase of FedShare}
\label{alg:fedshare_unlearning}
\begin{algorithmic}[1]
\Require Updated shared graph $\mathcal{G}_s$ after unsharing; unlearning requests $\{\mathcal{I}_u^{\mathrm{unlearn}}\}_{u\in\mathcal{U}}$; historical snapshots $\{\mathbf{V}_s^{(t-m)}\}_{m=1}^{M}$; current global parameters $\Theta_s=\{\mathbf{V}_s,\boldsymbol{\theta}_s\}$; unlearning rounds $T_{\mathrm{un}}$
\Ensure Unlearned global parameters $\tilde{\Theta}_s=\{\tilde{\mathbf{V}}_s,\boldsymbol{\theta}_s\}$
\State Construct forgotten graph $\mathcal{G}_{\mathrm{f}}$ using Eq.~\eqref{eq:forgotten_graph}
\For{$r=1,2,\ldots,T_{\mathrm{un}}$}
    \State Server samples a client subset $\mathcal{U}^{(r)} \subseteq \mathcal{U}$
    \State Server broadcasts $\{\mathbf{V}_s,\boldsymbol{\theta}_s\}$ to all $u\in\mathcal{U}^{(r)}$
    \ForAll{$u\in\mathcal{U}^{(r)}$ \textbf{in parallel}}
        \State Local training on remaining interactions (Sec.~\ref{sec:localtraining})
        \State Upload updated item embedding table $\mathbf{V}_u$ to server
    \EndFor
    \State Obtain new local-view embedding table $\mathbf{V}_{s,\mathrm{loc}}^{(r)}$ via FedAvg (Eq.~\eqref{eq:fedavg})
    \State Infer new global-view embeddings on updated $\mathcal{G}_s$ (Eq.~\eqref{eq:lgc_general}, \eqref{eq:layer_comb}) $\rightarrow \mathbf{V}_{s,\mathrm{glob}}^{(r)}$
    \For{$m=1,\ldots,M$}
        \State Infer forgotten-view embeddings on $\mathcal{G}_{\mathrm{f}}$ with $\mathbf{V}_s^{(t-m)}$ (Eq.~\eqref{eq:lgc_general}, \eqref{eq:layer_comb}) $\rightarrow \mathbf{V}_{\mathrm{f}}^{(t-m)}$
    \EndFor
    \State Update item embeddings by minimizing contrastive unlearning loss (Eq.~\eqref{eq:contrastive_unlearning}) to obtain $\tilde{\mathbf{V}}_s$
    \State Set $\mathbf{V}_s \leftarrow \tilde{\mathbf{V}}_s$
\EndFor
\end{algorithmic}
\end{algorithm}

\section{Experiment}\label{sec:experi}

To comprehensively evaluate the effectiveness of the proposed \textbf{FedShare} framework,
we design experiments to answer the following research questions:

\begin{itemize}
    \item \textbf{RQ1:} How does FedShare perform under personalized user data sharing compared with state-of-the-art federated recommender systems? 
    \item \textbf{RQ2:} In the unlearning setting, how does FedShare compare with state-of-the-art federated recommendation unlearning methods?
    \item \textbf{RQ3:} How do the major components of FedShare affect its performance?
    \item \textbf{RQ4:} How do different hyperparameter settings affect the performance of FedShare?
\end{itemize}

\subsection{Dataset}
We compare the proposed FedShare method with baseline methods on three
widely used public datasets.
\textbf{(1) MovieLens-1M\footnote{https://grouplens.org/datasets/movielens/}.}
MovieLens-1M is a movie rating dataset containing about one million user--movie ratings. Following common practice, we convert explicit ratings into implicit feedback, where a user is considered to have interacted with a movie if a rating exists. Users with fewer than 20 interactions are filtered out. \textbf{(2) Fashion \cite{ni2019justifying}.} This dataset consists of user reviews on fashion products from Amazon.
All reviews are treated as implicit user--item interactions.
Similarly, users with fewer than 20 interactions are removed. \textbf{(3) Video Game \cite{ni2019justifying}.}
This dataset contains user reviews on video game products from Amazon.
We also convert all reviews into implicit feedback and filter out users with fewer than
10 interactions.

For each dataset, we randomly split the interactions into training, validation, and test sets with a ratio of $8{:}1{:}1$.
During the data sharing phase, users are divided into three groups with a ratio of $1{:}2{:}7$, corresponding to fully sharing users, partially sharing users (with a sharing
ratio of $0.3$), and non-sharing users, respectively.
The impact of different user sharing ratios is further analyzed in
Section~\ref{sec:hyper}. In the unlearning phase, we set the unsharing ratio to $0.3$, meaning that $30\%$ of the
users who previously shared data request to remove their shared interactions and the
corresponding influence on the trained model.
We also study the effect of different unsharing ratios in
Section~\ref{sec:hyper}. Overall statistics of the three datasets are summarized in Table~\ref{tab:datasets}.

\begin{table}[!htbp]
  \centering
  \caption{Statistics of datasets}
    \begin{tabular}{c|c|c|c}
    \toprule
    Dataset & \# Users & \# Items & \# Interactions \\
    \midrule
    \midrule
    MovieLens-1M & 6,040 & 3,900 & 1,000,209  \\
    Fashion & 29,858 & 40,981 & 1,027,370  \\
    Video Games & 15,517 & 37,077 & 284,867 \\
    \bottomrule
    \end{tabular}
\label{tab:datasets}
\end{table}

\subsection{Baseline}

We compare the proposed \textbf{FedShare} with two categories of state-of-the-art methods:
(i) federated recommender systems with personalized user data sharing (\emph{Learning methods}),
and (ii) federated recommendation unlearning methods (\emph{Unlearning methods}).

\textbf{Learning Methods:}
\begin{itemize}
    \item \textbf{FedShare-FedAvg.}
    A straightforward baseline under the personalized user data sharing setting. The server is treated as a special client that holds all user-shared interactions, and standard FedAvg is applied to perform federated training.

    \item \textbf{CDCGNNFed}~\cite{qu2024towards}.
    Our previous work that addresses incomplete shared data by introducing a graph mending strategy on the server side to predict potentially missing user--item interactions. It further employs client--server collaborative contrastive learning to refine user and item embeddings.

    \item \textbf{PDC-FRS}~\cite{yang2024pdc}.
    A federated recommendation framework with a privacy-preserving data contribution mechanism, where differential privacy techniques are used to protect user-shared data while enabling
    collaborative model training.
\end{itemize}

\textbf{Unlearning Methods:}
\begin{itemize}
    \item \textbf{Retrain.}
    A full retraining strategy that retrains the model from scratch using only the remaining data after unlearning requests.
    Although computationally expensive, it serves as the ground-truth reference for evaluating unlearning effectiveness.

    \item \textbf{FRU}~\cite{yuan2023federated}.
    A federated recommendation unlearning method that maintains important historical gradients on the client side. After unlearning requests are issued, gradient correction is applied to remove the influence of the forgotten data.

    \item \textbf{CUFRU}~\cite{li2025cross}.
    A server-side federated recommendation unlearning method that introduces a gradient transfer station. CUFRU leverages both historical gradients and newly updated gradients from remaining clients to perform gradient calibration.
\end{itemize}

\subsection{Evaluation metrics}
We adopt two commonly used metrics in recommender systems to evaluate recommendation
performance, including Hit Ratio (HR@K) and Normalized Discounted Cumulative Gain
(NDCG@K).
HR@K reflects whether the ground-truth item is successfully recommended within the
top-$K$ list, while NDCG@K evaluates the ranking quality by assigning higher importance to
relevant items that appear earlier in the recommendation list.

\subsection{Experimental Setup}
For all methods, we set the embedding dimension to $d=32$ and initialize all embedding
tables using the Xavier initialization \cite{glorot2010understanding}.
For the LGC-based recommender, we use $L=3$ propagation layers on the server-side shared
high-order user--item graph.
For both the learning and unlearning phases, the temperature parameter $\tau$ in
contrastive learning and contrastive unlearning is selected from
$\tau \in \{0.05, 0.1, 0.2, 0.5, 1.0\}$ based on validation performance.
For the learning phase, the weight of the contrastive learning loss $\lambda_1$ is tuned
from $\{0.1, 0.3, 0.5, 0.7\}$.
For the unlearning phase, the number of historical global embedding tables $M$ is chosen
from $\{1, 2, 3, 5, 10\}$.

\subsection{Experimental results}

\subsubsection{Recommendation Performance in Learning and Unlearning Phases (RQ1, RQ2)}
\begin{table*}[htbp]
  \centering
  \caption{Recommendation performance comparison under personalized data sharing (Learning phase) and federated unlearning (Unlearning phase).}
    \begin{tabular}{c|l|cc|cc|cc}
    \toprule
    \multirow{2}[4]{*}{Phase} & \multicolumn{1}{c|}{\multirow{2}[4]{*}{Method}} & \multicolumn{2}{c|}{MovieLens-1M} & \multicolumn{2}{c|}{Fashion} & \multicolumn{2}{c}{Video Game} \\
\cmidrule{3-8}          &       & HR@20 & NDCG@20 & HR@20 & NDCG@20 & HR@20 & NDCG@20 \\
    \midrule
    \midrule
    \multirow{4}[2]{*}{Learning} & FedShare-FedAvg & 0.4656 & 0.0642 & 0.0564 & 0.0053 & 0.2533 & 0.0174 \\
          & PDC-FRS & 0.4662 & 0.0647 & 0.0569 & 0.0057 & 0.2576 & 0.0176 \\
          & CDCGNNFed & 0.4685 & 0.0651 & 0.0588 & 0.0069 & 0.2532 & 0.0185 \\
          & FedShare & \textbf{0.4773} & \textbf{0.0659} & \textbf{0.0607} & \textbf{0.0074} & \textbf{0.2614} & \textbf{0.0198} \\
    \midrule
    \midrule
    \multirow{4}[2]{*}{Unlearning} & Retrain & 0.4682 & 0.0651 & 0.0552 & 0.0063 & 0.2534 & 0.0191 \\
          & FRU   & 0.4632 & 0.0634 & 0.0531 & 0.0046 & 0.2522 & 0.0163 \\
          & CUFRU & 0.4634 & 0.0633 & 0.0526 & 0.0048 & 0.2515 & 0.0159 \\
          & FedShare & \textbf{0.4665} & \textbf{0.0647} & \textbf{0.0549} & \textbf{0.0056} & \textbf{0.2526} & \textbf{0.0172} \\
    \bottomrule
    \end{tabular}%
  \label{tab:TOPK}%
\end{table*}%

\begin{figure*}
    \centering
    \includegraphics[width=0.8\linewidth]{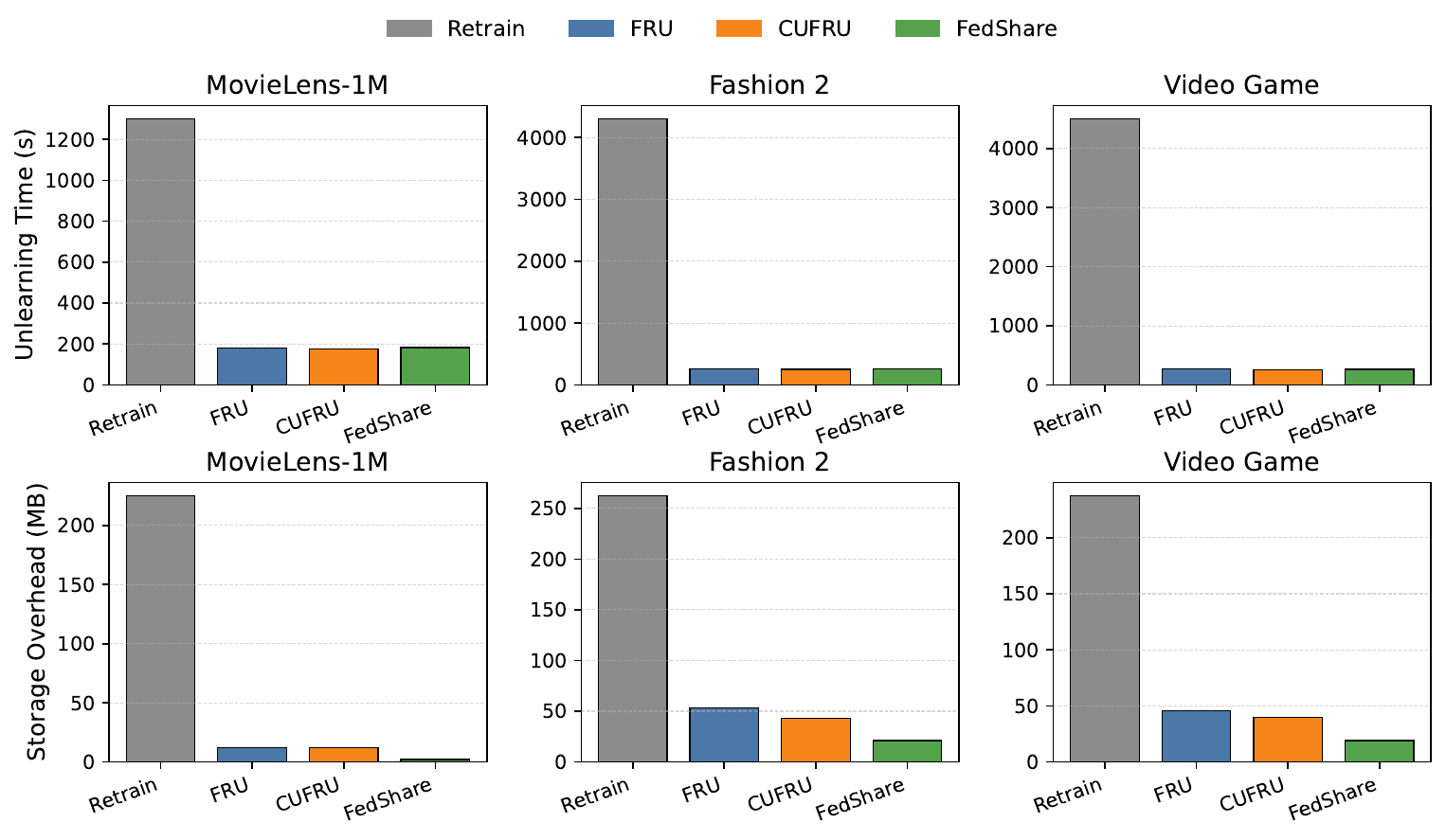}
    \caption{Comparison of unlearning efficiency and storage overhead of different federated recommendation unlearning methods on three datasets.}
    \label{fig:unlearningeff}
\end{figure*}
We first evaluate the recommendation performance of different methods on the Top-$K$
recommendation task under both the learning phase and the unlearning phase.
The results are reported in Table~\ref{tab:TOPK}.
From the results, we obtain the following observations. (1)Under the personalized user data sharing setting, methods that explicitly exploit shared interaction data, including PDC-FRS, CDCGNNFed, and FedShare, consistently
outperform the naive baseline that simply treats the server as a special client
(i.e., FedShare-FedAvg) in most cases.
This demonstrates the necessity of our research motivation: allowing users to share data
in a personalized manner enables the server to leverage additional collaborative signals,
leading to improved recommendation performance.
(2) FedShare achieves the best overall performance in the learning phase across all
three datasets. We attribute this improvement to the introduction of contrastive learning, which
effectively strengthens item representations by aligning local and global views.
This design allows the model to jointly capture low-order collaborative signals from
client-side interactions and high-order structural information from the server-side
shared graph. (3) In the unlearning phase, FedShare consistently outperforms existing federated recommendation unlearning methods, including FRU and CUFRU, in terms of recommendation accuracy. A possible explanation is that the proposed contrastive unlearning strategy selectively removes representation components induced by the unshared data, while preserving representations supported by the remaining data. As a result, FedShare avoids the performance degradation commonly observed in gradient-calibration-based unlearning methods. (4) Although FedShare does not outperform the retrain-from-scratch oracle in terms of recommendation accuracy, this result is expected. The retrain method completely remove the influence of forgotten data by fully retraining the model on the remaining data. In contrast, FedShare is designed as an efficient unlearning solution that avoids costly full retraining.

\begin{table*}[htbp]
  \centering
  \caption{Ablation study results of FedShare on three datasets.}
    \begin{tabular}{l|cc|cc|cc}
    \toprule
    \multicolumn{1}{c|}{\multirow{2}[2]{*}{Variant}} & \multicolumn{2}{c|}{MovieLens-1M} & \multicolumn{2}{c|}{Fashion} & \multicolumn{2}{c}{Video Game} \\
          & HR@20 & NDCG@20 & HR@20 & NDCG@20 & HR@20 & NDCG@20 \\
    \midrule
    \midrule
    FedShare (Learing) & 0.4773 & 0.0659 & 0.0607 & 0.0074 & 0.2614 & 0.0198 \\
    \midrule
    w/o CL & 0.4702 & 0.0634 & 0.0597 & 0.0072 & 0.2584 & 0.0183 \\
    w/o BPR & 0.4725 & 0.0638 & 0.0593 & 0.0066 & 0.2576 & 0.018 \\
    \midrule
    FedShare (UnLearing) & 0.4665 & 0.0647 & 0.0549 & 0.0056 & 0.2526 & 0.0172 \\
    \midrule
    w/o forgotten graph & 0.4624 & 0.0613 & 0.0524 & 0.0049 & 0.2513 & 0.0164 \\
    w/o remaining FL & 0.4642 & 0.0635 & 0.0533 & 0.0051 & 0.2519 & 0.0166 \\
    \bottomrule
    \end{tabular}%
  \label{tab:ablation}%
\end{table*}%

\begin{figure*}
    \centering
    \includegraphics[width=1\linewidth]{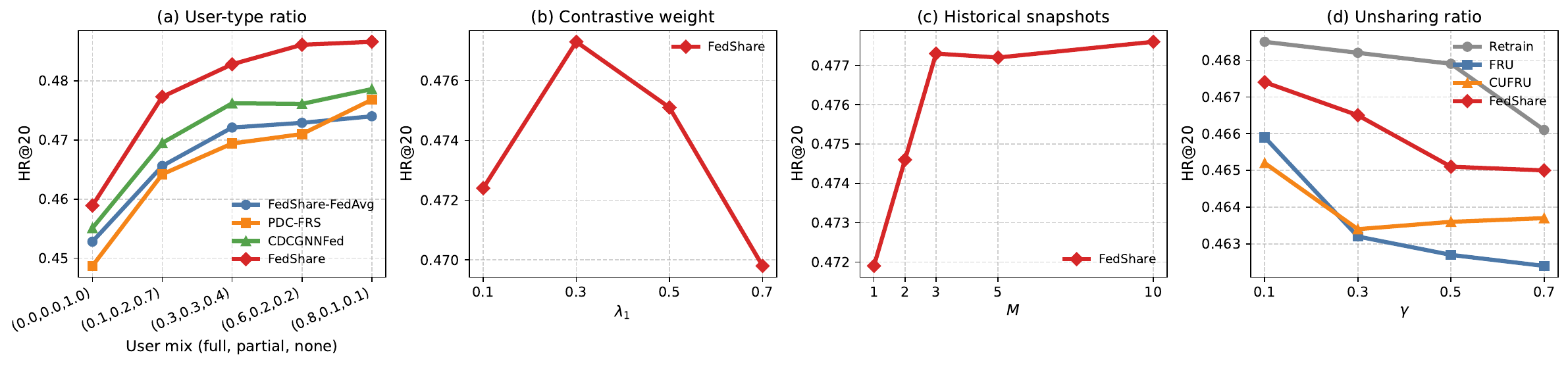}
    \caption{Results of hyperparameter analysis on (a) user-type ratio, (b) contrastive weight, (c) number of historical snapshots, and (d) unlearning ratio.}
    \label{fig:hyperparam}
\end{figure*}

\subsubsection{Unlearning Efficiency and Storage Overhead (RQ2)}
We further compare different federated recommendation unlearning methods in terms of
unlearning efficiency and storage overhead.
The experimental results are reported in Fig.~\ref{fig:unlearningeff}. From the results, we make the following observations. (1) FedShare achieves competitive unlearning efficiency compared with FRU and CUFRU, and is significantly more efficient than the retrain-from-scratch baseline. This behavior is expected, as retraining requires repeating the entire federated learning process on the remaining data.
In contrast, FedShare updates the model through targeted contrastive unlearning, which
efficiently removes the influence of unshared data without full retraining.
(2) FedShare shows lower storage overhead than FRU and CUFRU across all datasets. This advantage stems from the fact that FedShare does not rely on storing large amounts of historical gradient information for gradient calibration. Instead, it only maintains a small number of historical item embedding tables to construct
forgotten views for contrastive unlearning, resulting in significantly reduced storage requirements.

\subsubsection{Ablation study (RQ3)}
To examine the contribution of each component in FedShare, we conduct ablation
studies by selectively removing key modules in the learning and unlearning phases.
In the learning phase, we consider \textit{w/o CL}, which removes the contrastive learning
module and relies only on FedAvg aggregation, and \textit{w/o BPR}, which excludes the BPR
loss during server-side refinement and optimizes the model using contrastive learning alone.
In the unlearning phase, we evaluate \textit{w/o forgotten graph}, where forgotten views
are constructed solely from historical embedding snapshots without using the forgotten
graph, and \textit{w/o remaining FL}, which skips federated training on the remaining data and directly reuses the previous local views. 
We report the ablation study results in Table~\ref{tab:ablation}. We can observe that:
(1) Removing the contrastive learning module in the learning phase leads to a significant performance drop, which confirms the necessity of aligning local and global views via contrastive learning to enhance representation quality.
(2) Excluding the BPR loss during global refinement also degrades performance. This is expected, since the shared interactions are not involved in local training; without the BPR objective, the model lacks direct supervision to capture high-order collaborative signals from the shared graph.
(3) In the unlearning phase, removing the forgotten graph leads to a performance drop.
This shows that explicitly modeling the forgotten interactions is important; without this information, the model may remove too much information and negatively affect representations learned from the remaining data.
(4) Using only historical local views without re-running federated training on the remaining clients further reduces performance, as the resulting representations fail to accurately reflect the updated data distribution after data unsharing.

\subsubsection{Hyper-parameter Analysis (RQ4)}\label{sec:hyper}
We further analyze the sensitivity of FedShare to several key hyper-parameters.
All hyper-parameter studies are conducted on MovieLens-1M dataset, while keeping
all other settings fixed. We evaluate the performance using HR@20. Specifically, we investigate:
(i) the ratio of different user types (fully-sharing, partially-sharing, and non-sharing users),
(ii) the contrastive learning weight $\lambda_1$ in the learning phase,
(iii) the number of historical embedding snapshots $M$ used to construct forgotten views,
and (iv) the unsharing ratio $\gamma$, which controls the scale of unlearning requests.
The results are summarized in Fig.~\ref{fig:hyperparam}.
We can observe that: (1) As the proportion of users who share data increases, the recommendation performance of all methods consistently improves. This is expected, since more shared interactions enable the server to construct a richer high-order user–item graph, allowing the model to better leverage collaborative signals across users.
(2) When the contrastive learning weight is either too small or too large, the model performance drops noticeably. This indicates that an appropriate balance between the BPR supervision and the contrastive objective is necessary; otherwise, the model fails to effectively capture collaborative information.
(3) As the number of historical snapshots 
M increases, the performance improves and gradually stabilizes. This is reasonable because more historical embeddings provide a more complete representation of the influence induced by forgotten interactions, while excessive snapshots bring diminishing returns.
(4) With a larger unlearning ratio, the performance of all methods decreases, as more information is removed from the model. Nevertheless, FedShare consistently outperforms the baseline methods under different unlearning ratios, demonstrating its robustness in handling varying unlearning demands

\section{Conclusion}\label{sec:con}
In this work, we introduce a fully personalized user data sharing framework for federated recommender systems.
Different from traditional federated recommender systems that adopt a one-size-fits-all data management strategy, our framework allows users to flexibly control how much interaction data is shared with the server to improve recommendation performance. More importantly, it also allows users to later request the removal of previously shared data together with its influence on the trained model. This setting fills an important gap in existing federated recommender systems, where personalized data sharing and data unlearning are rarely considered in a unified manner.
To support this framework, we propose a contrastive learning module to better utilize the shared user–item graph during the learning phase, enabling the model to effectively capture both low-order and high-order collaborative signals. In addition, we design a contrastive unlearning mechanism that selectively removes the influence of unshared data by leveraging a small number of historical embedding tables, rather than storing large amounts of historical gradient information as required by existing federated recommendation unlearning methods.
Extensive experiments on three public datasets demonstrate that FedShare can effectively leverage shared data and achieve better recommendation performance in the learning phase compared with state-of-the-art baselines. In the unlearning phase, our method significantly reduces storage requirements while maintaining competitive recommendation accuracy.

Despite its effectiveness, this work has two main limitations.
First, when handling data unlearning requests, the proposed framework still requires the participation of remaining users to assist the unlearning process, which may increase client-side communication and computation overhead. In future work, we plan to explore fully server-side unlearning mechanisms to reduce the need for additional client participation.
Second, the current framework focuses on users who have previously shared data with the server and later request the removal of their data and its corresponding influence. It does not yet support unlearning requests from users who never shared data but only participated in federated training. Extending the framework to support such users is an important direction for future research.

\bibliographystyle{IEEEtran}
\bibliography{main}

\end{document}